\documentclass[11pt]{article} 
\usepackage[colorlinks=true,urlcolor=blue,citecolor=blue]{hyperref} 
\pdfoutput=1 

\usepackage[numbers]{natbib}

\begin{document} 
\title{Mass Transfer in Bubbly Flow Using \\ a Subscale Description} 
\author{Bahman Aboulhasanzadeh, Gr\'etar Tryggvason \\ 
\\\vspace{6pt} Department of Aerospace and Mechanical Engineering, \\ University of Notre Dame
Notre Dame, IN  46556-5637, USA} 
\maketitle 
\thispagestyle{empty}
\begin{abstract} 
In the computation of multiphase flow with mass transfer, the large disparity between the length and time scale of the mass transfer and the fluid flow demand excessive grid resolution for fully resolved simulation of such flow.  We have developed a subscale description for the mass transfer in bubbly flow to alleviate the grid requirement needed at the interface where the mass gets transferred from one side to the other.

In this fluid dynamics video, a simulation of the mass transfer from buoyant bubbles is done using a Front Tracking method for the tracking of interface and a subscale description for the transfer of mass from the bubble into the domain. After the mass is transferred from the bubble into the domain, mass is followed by solving an advection-diffusion equation on a relatively coarse Cartesian grid. More detail about the method can be found in our \href{http://dx.doi.org/10.1016/j.ces.2012.04.005}{paper}[\citenum{Aboulhasanzadeh:2012aa}].

This simulation shows 13 moving bubbles in a periodic domain, $3d_b\times 3d_b \times 48d_b$, where $d_b$ is the bubble diameter. The grid resolution is $64\times 64 \times 1024$, which results in about 21 cell across one bubble diameter. The flow non-dimensional governing parameters are $Eo = 2.81$ and $Mo = 4.5\times 10^{-7}$ with density and viscosity ratio of $0.1$ and for the mass transfer we have $Sc=60$. In the movie, bubbles are colored to show the mass boundary layer thickness, with blue showing a close to zero value and red showing the maximum value. Mass concentration inside the domain is colored from transparent blue for low value, 0, to solid red for high value, 1. Time is non-dimensionalized with $\sqrt{d_b/g}$.\\

\end{abstract} 

\bibliographystyle{elsarticle-harv}

\end{document}